\documentclass[aps,prl,reprint,bibnotes,superscriptaddress,nofootinbib]{revtex4-2}
\usepackage{graphicx,amsmath}
\usepackage{xcolor}
\usepackage{tablefootnote}
\begin{document}

%\preprint{}

\title{Sub-Doppler laser cooling and optical transport of cesium with static magnetic fields}



\author{Tobias Bothwell}
\email{tobias.bothwell@infleqtion.com}
\affiliation{Infleqtion, Boulder, Colorado}

\author{Junxin Chen}
\affiliation{Infleqtion, Boulder, Colorado}

\author{Brian M. Fields}
\affiliation{Infleqtion, Boulder, Colorado}

\author{Madeline K. Dawes}
\affiliation{Infleqtion, Boulder, Colorado}

\author{Anthony Reiter}
\affiliation{Infleqtion, Boulder, Colorado}

\author{Christina C. C. Willis}
\affiliation{Infleqtion, Boulder, Colorado}

\author{Jacob Scott}
\affiliation{Infleqtion, Madison, Wisconsin}

\author{Michael McMaster}
\affiliation{Infleqtion, Boulder, Colorado}

\author{Farhad Majdeteimouri}
\affiliation{Infleqtion, Boulder, Colorado}

\author{Ilya Vinogradov}
\affiliation{Infleqtion, Boulder, Colorado}

\author{Seth Miers}
\affiliation{Infleqtion, Boulder, Colorado}

\author{Daniel C. Cole}
\affiliation{Infleqtion, Boulder, Colorado}

\author{Kevin Loeffler}
\affiliation{Infleqtion, Boulder, Colorado}

\author{Ryan A. Jones}
\affiliation{Infleqtion, Boulder, Colorado}

\author{Marin Iliev}
\affiliation{Infleqtion, Boulder, Colorado}

\author{Jonathan Gilbert}
\affiliation{Infleqtion, Boulder, Colorado}

\author{Eric Copenhaver}
\affiliation{Infleqtion, Boulder, Colorado}

\author{Thomas W. Noel}
\affiliation{Infleqtion, Boulder, Colorado}

\author{Alexander G. Radnaev}
\affiliation{Infleqtion, Boulder, Colorado}
%\email{\textsuperscript{$\dagger$}andrew.ludlow@nist.gov}
\date{\today}

\begin{abstract}

Laser cooling of alkali atoms typically requires time-varying magnetic fields, introducing unwanted coupling between atom preparation and coherent operations. Here we demonstrate sub-Doppler laser cooling and optical transport of alkali atoms in a fully static magnetic-field configuration. Using a blue-detuned Type-II magneto-optical trap (MOT) operating on the closed $F=3 \rightarrow F'=2$ transition of the D2 line in cesium, we achieve temperatures of 17(1) µK without changing the magnetic-field gradient between cooling stages. This enables direct loading into a shallow optical lattice and transport over 17 cm within the same static-field environment. In contrast to conventional alkali cooling schemes with dynamic fields, our approach establishes a continuous cooling and transport protocol compatible with static-field platforms. These results validate Type-II cooling as a practical technique for alkali atoms and provide a new route toward continuous-operation architectures in sensing and quantum computing.

\end{abstract}

%\keywords{}

\maketitle

%\footnote{\label{a}$^*$ These authors contributed equally}
%\footnote{\label{b}$^{\dag}$ chenchunchia@gmail.com, andrew.ludlow@nist.gov}

\textit{Introduction}--Continuously operating atomic platforms enable new capabilities in quantum sensing~\cite{dick1989local,holland1996theory,meiser2009prospects,kwolek2022continuous} and computation~\cite{aharonov1997fault,kitaev1997quantum,li2024high} by spatially decoupling atom preparation from regions requiring long coherence times. This paradigm has enabled advances ranging from steady-state Bose–Einstein condensation in strontium~\cite{chen2022continuous} to large-scale, defect-free neutral-atom arrays with continuous atom resupply~\cite{norcia2024iterative,chiu2025continuous}. As such architectures mature, reducing the complexity of atom preparation becomes increasingly important. In alkali-based systems, sub-Doppler cooling typically relies on time-varying magnetic fields~\cite{ketterle1999making,metcalf1999laser}, which may introduce unwanted perturbations to nearby coherent qubits. In contrast, narrow-line cooling in alkaline-earth-like atoms can be performed in static magnetic fields, naturally avoiding these complications~\cite{norcia2024iterative}. This highlights a central challenge: realizing sub-Doppler laser cooling and optical trapping of alkali atoms in static magnetic fields.

Over the past decade, blue-detuned MOTs (BDMs) based on  Type-II transitions have emerged as a powerful approach to sub-Doppler cooling~\cite{jarvis2018blue}. Unlike standard Type-I MOTs, which operate on transitions with ground (excited) state total angular momentum $F$ ($F'=F+1$), Type-II MOTs leverage transitions between $F\rightarrow F'\leq F$, enabling polarization-dependent dark states and sub-Doppler cooling mechanisms that are robust to the magnetic field of the MOT~\cite{tarbutt2015magneto,devlin2016three}. The first Type-II BDM was demonstrated in Rb~\cite{jarvis2018blue}, and the technique has since been applied to molecular laser cooling of YO~\cite{burau2023blue} and CaF~\cite{li2024blue}. However, the use of the Type-II BDMs  has remained largely limited to molecular species despite the apparent advantages of narrow-line cooling behavior.

In this Letter, we demonstrate sub-Doppler laser cooling and optical transport of cesium in a static magnetic field configuration using a Type-II BDM. By operating both Type-I and Type-II MOT stages with an identical magnetic field gradient, we achieve this sub-Doppler cooling without magnetic field switching. We load the resulting cold ensemble into a one-dimensional optical lattice and transport millions of atoms over 17 cm while maintaining a fixed magnetic field gradient centered at the loading site. These results establish a new route to static-field-compatible laser cooling in alkali systems and enable simplified architectures for continuously operated neutral-atom platforms.

\textit{Experimental Overview}--Our experiments occur in a dual-cell vacuum chamber composed of a source cell with Cs dispenser and science cell~(Fig.~1a). To spatially separate initial laser cooling from qubit arrays in the science cell, we form a MOT in the source cell using a free-space optical setup with three pairs of centimeter-diameter retro-reflected MOT beams. A standard red-detuned Type-I MOT (Fig.~1b)~\cite{monroe1990very} provides initial laser cooling from vapor and trapping of Cs in a quadrupole magnetic field of $B_{MOT}=17$~G/cm along the axial axis of the coils. Repump light addressing $F=3\rightarrow F'=4$ on the D2 line is overlapped with two MOT arms to return atoms lost from the cooling cycle. We achieve Type-I MOT atom numbers of approximately 50 million.

\begin{figure*}[t]
    \includegraphics[trim={0.5cm 0.6cm 0 0.8cm},clip,width=.9\textwidth]{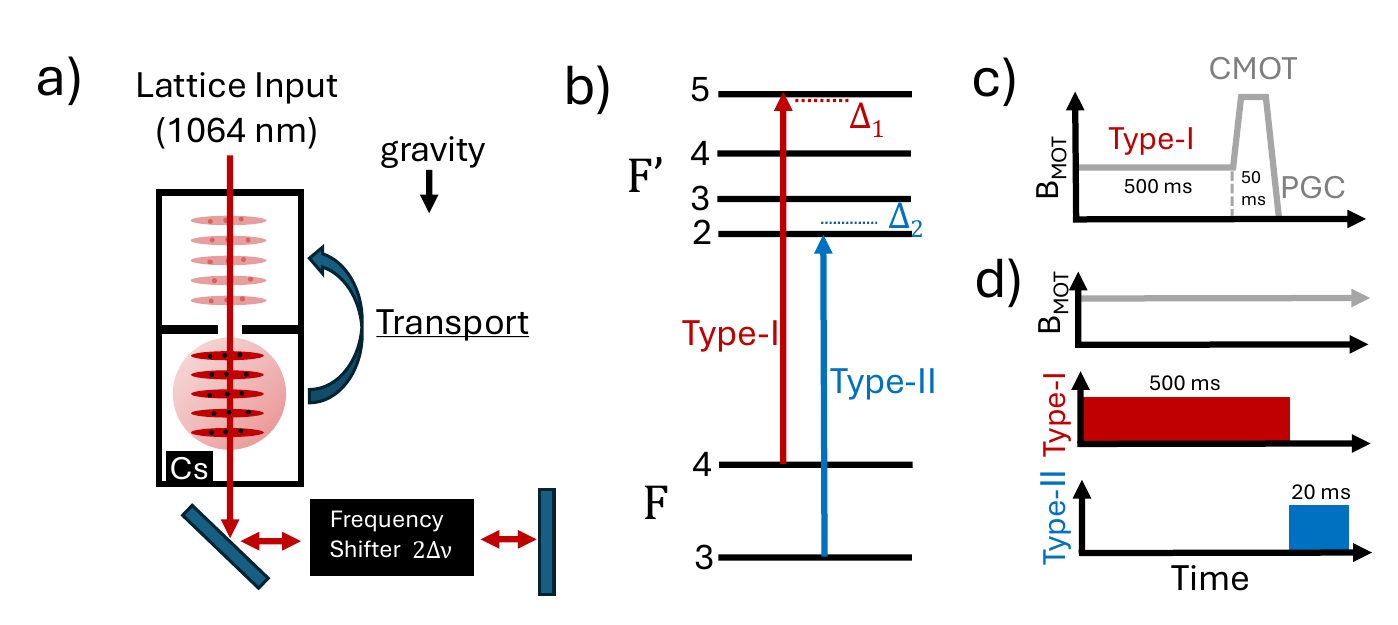}
    \vspace{-8pt}

    \caption{\label{figure_1}
    (a) Experimental apparatus. A dual-cell vacuum chamber with Cs background vapor in a source cell is separated from a high-vacuum science cell via differential pumping aperture. We form a MOT (red circle) directly from the background Cs vapor which provides cold atoms trapped in a 1D optical lattice. We control a variable frequency offset on the retro-reflected beam to optically transport atoms to the science cell 17 cm away.
    (b) Hyperfine states of the D2 transition in Cs, with total angular momentum $F$ ($F'$) shown for the ground (excited) 6$S_{1/2}$ (6$P_{3/2}$) state. The Type-I MOT operates with red detuning ($\Delta_1=-8(3)$ MHz). Cs has a closed Type-II transition on the $F=3\rightarrow F'=2$ transition, which supports sub-Doppler cooling when blue-detuned ($\Delta_2=25(3)$ MHz). Leakage from off-resonant scatter from nearby hyperfine levels is addressed via repumping (see text for details).
    (c) Standard loading of alkali atoms into a 1D optical lattice may employ a combination of compression MOT (CMOT) and polarization-gradient cooling (PGC), requiring time-varying magnetic fields.
    (d) Type-I and subsequent Type-II MOTs eliminate the requirement for varying magnetic fields for sub-Doppler cooling.}
\end{figure*}

To transport atoms between vacuum chamber cells we employ a red-detuned 1D optical lattice where the atoms are trapped at anti-nodes of the standing wave (Fig.~1a)~\cite{grimm2000optical}. The lattice focuses 8 W of 1064 nm light to a waist of $\omega_0=232(6)$~µm at the MOT and then passes through a pair of acousto-optic modulators (AOMs) before getting retroreflected, allowing for a variable frequency offset ($2\Delta \nu$) between the beams. This first traps atoms in a stationary standing wave before chirping the frequency offset to provide controllable acceleration for optical transport~\cite{klostermann2022fast}. Our lattice has a trap depth of 51(7)~µK at the MOT as measured by parametric heating~\cite{jauregui2001anharmonic}. At these depths standard MOT operation is unable to provide efficient loading of the transport lattice, requiring additional laser cooling stages. A deeper optical potential could be employed, but sub-Doppler cooling would still be required for subsequent transport and optical tweezer loading~\cite{chiu2025continuous}. Techniques such as compression MOTs~\cite{petrich1994behavior} and polarization gradient cooling (Fig.~1c)~\cite{dalibard1989laser} are standard solutions to this problem, but they require time-varying magnetic fields which couple to qubits in the science cell.

 \begin{figure}[b]
	\includegraphics[trim={0 0.0cm 0 0.0cm},clip,width=.49\textwidth]{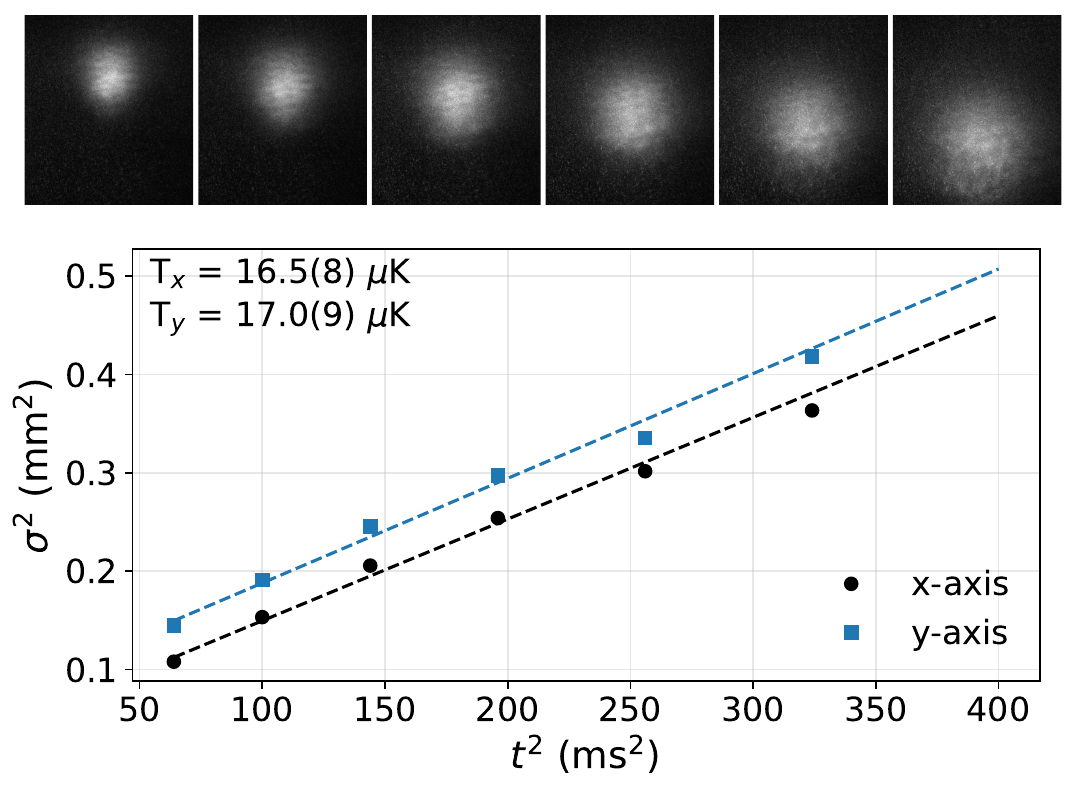}
    \vspace{-20pt}
	\caption{\label{1}\emph{Top} Images of time-of-flight (TOF) showing atomic cloud drop and ballistic expansion. \emph{Bottom} Quadratic cloud width ($\sigma^2$) versus quadratic TOF time ($t^2$). x (y) corresponds to perpendicular (parallel) axis to gravity. Error bars from statistical uncertainty from fitting a 2D Gaussian to each image are too small to see.}
	\label{figure_2}
\end{figure}

\begin{figure*}[!t]
    \includegraphics[trim={0 0.0cm 0 0.0cm},clip,width=.9\textwidth]{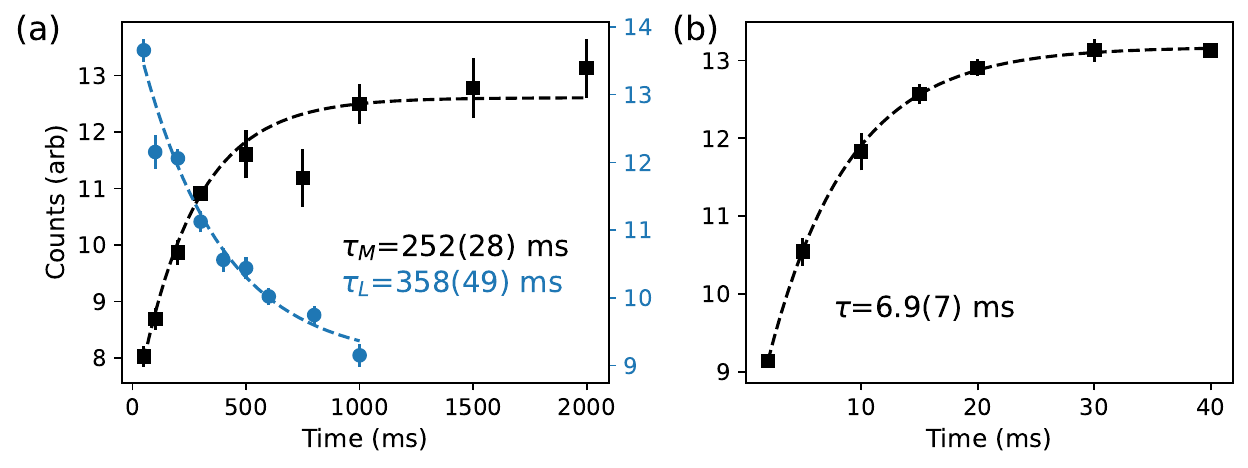}
    \vspace{-10pt}
    \caption{\label{figure_3}
    (a) Exponential saturation (decay) fits for the number of atoms in the lattice after varying MOT loading time (dark lattice hold time) in black (blue). Arbitrary counts are shown due to different loading conditions and gain settings used for imaging. The lattice lifetime suggests a background-gas-limited vacuum lifetime of $\tau_L=358(49)$ ms, with the MOT loading curve providing a lower bound of $\tau_M=252(28)$ ms likely reduced due to imperfect alignment and repumping.
    (b) Lattice signal for varying BDM durations. The time constant for BDM performance is far below standard Type-I MOT loading times and comparable to lattice transport times, providing sub-Doppler cooling that supports high shot rates for atom transport.}
\end{figure*}

To support high-fidelity qubit operations with concurrent atom preparation, we design a new static-magnetic field atomic preparation scheme. Following the initial Type-I MOT, we employ a BDM operating on the closed $F=3 \rightarrow F'=2$ transition of the D2 line (Fig. 1b). In contrast to Rb which requires simultaneous operation of two BDMs~\cite{jarvis2018blue}, cesium's hyperfine structure enables BDM operation with a single additional MOT frequency. The Cs BDM operates with opposite helicity of the Type-I MOT beams~\cite{devlin2016three}, with the 5 mm diameter BDM beams readily overlapped at high efficiency with our free space Type-I MOT system (End Matter). While closed, the Type-II MOT leaks atoms due to off-resonant scatter which we repump via $F=4 \rightarrow F'=4$ light on the D1 line with a single auxiliary beam, chosen for hardware compatibility. Throughout this paper we operate with BDM parameters of detuning $\Delta_2=$ 25(3) MHz, beam diameters of 4.9(5) mm, and intensity of 18(4) mW/cm$^2$ for the BDM beams. We operate with a 20 ms BDM stage and utilize the same magnetic field gradients as the Type-I MOT (Fig. 1d).

\textit{Laser Cooling and Trapping Performance}--Type-II BDM performance across a wide array of parameter regimes has been studied in Rb~\cite{jarvis2018blue}. Motivated by the robustness of the reported BDM, we optimized our two-stage MOT operation for loading of our optical lattice. Once we settled on the optimal $B_{MOT}$ for lattice loading performance, we returned to the BDM to evaluate the temperature via time-of-flight (TOF) imaging~\cite{ketterle1999making}. The TOF experiments are the only ones in this Letter where we change a magnetic field and operate with the optical lattice off. We turn off the quadrupole field after the BDM is complete, allowing the atomic cloud to fall and undergo ballistic expansion. Fig. 2 shows the images collected for varying TOF times, with each fit to a 2D Gaussian to evaluate initial cloud size and temperature. Pixel size is calibrated by the center-of-mass drop from gravity, allowing us to extract the effective pixel size of our imaging system and temperatures for the cloud. We find initial widths of approximately 250 µm and temperatures of 17(1)~µK,  far below the Doppler cooling limit of 120~µK.

Next we study laser cooling and optical trapping dynamics within the source cell. Throughout the remainder of this Letter magnetic fields are static and the 1D optical lattice is always on with a constant depth of $U=51(7)$~µK. With a cold atomic sample from the BDM trapped in the lattice we can extinguish the MOT beams, wait 50 ms for untrapped background atoms to disperse, and then image the atoms in the optical lattice with the Type-I MOT beams. A critical figure of merit for loading dynamics is the background-gas-limited lifetime ($\tau_g$) of the atoms in the source cell. We operate at a reduced dispenser current compared with standard 2D MOT operation, enabling background-gas-limited lifetimes $>100$ ms in the source cell at the expense of MOT loading rates.  We probe $\tau_g$ via two methods: 1) MOT loading and 2) lattice lifetime (Fig.~3a). For low atomic densities and assuming closed Type-I MOT operation, the dynamics of atom number $(N)$ versus Type-I MOT loading time ($t_{MOT}$) can be modeled by an exponential saturation curve ($N(t)=N_{max}(1-\exp{[-t/\tau_M]})$)~\cite{arpornthip2012vacuum}. To measure $\tau_M$ we image atoms trapped in the optical lattice, varying the Type-I MOT loading duration, and find $\tau_M$=252(28) ms, a lower bound on $\tau_g$. We confirm a longer lifetime via lattice lifetime measurements. Holding the Type-I and Type-II MOT durations constant, we vary the dark hold time in the lattice before imaging the remaining atoms. Fitting an exponential decay to this process we find $\tau_L$=358(49) ms, slightly longer than the MOT lifetime. Both are sufficient for loading and optical transport timescales.

\begin{figure*}[t]
    \includegraphics[trim={0 0.0cm 0 0.0cm},clip,width=.95\textwidth]{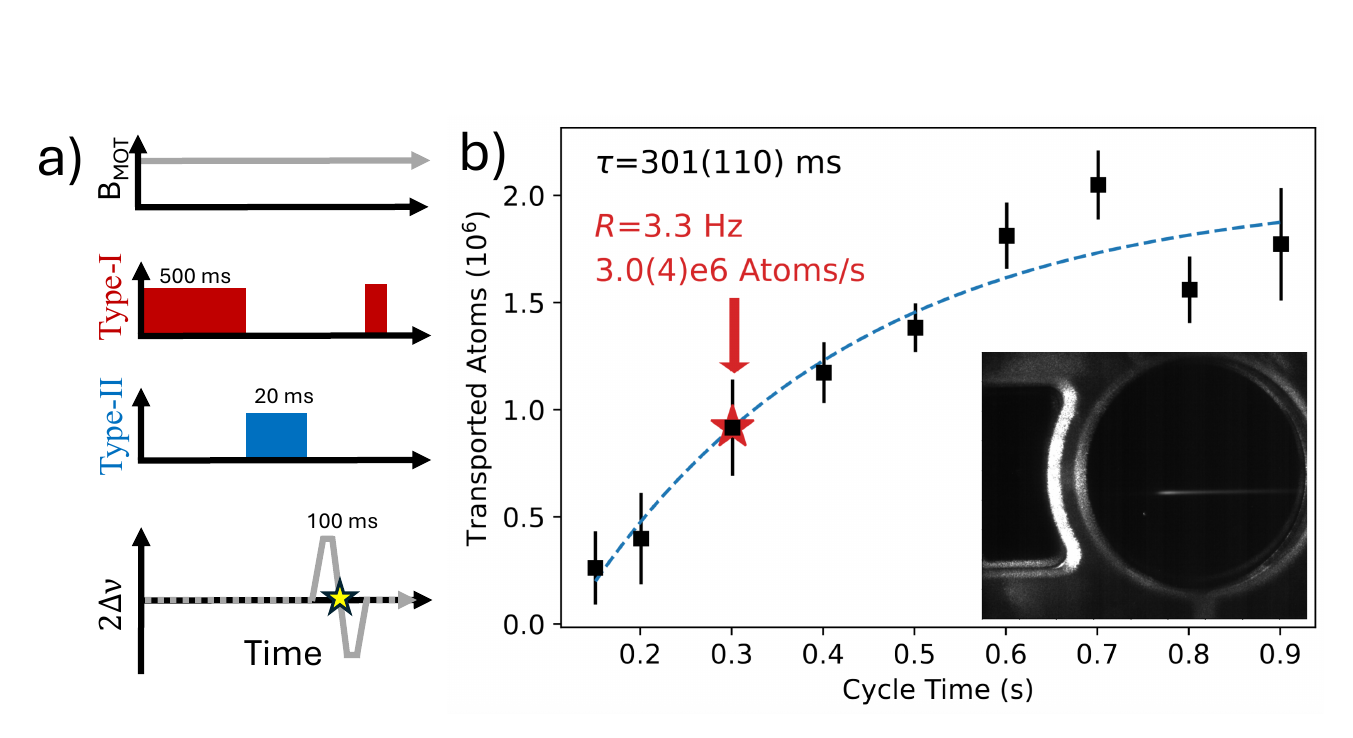}
    \vspace{-8pt}
    \caption{\label{figure_4}
    (a) Timing diagrams as in Fig.~\ref{figure_1}. The lattice frequency offset is ramped, moving atoms to the science chamber (yellow star) before ramping back to the source cell to be imaged with the Type-I system.
    (b) Exponential saturation curve of round-trip transported atom number. The time constant is consistent with the source-cell lifetime, showing transported atom number is dictated by MOT loading dynamics. Optimal transport flux of $3.0(4)\times10^6$ atoms/s (red star) is obtained when cycle time and saturation time constant match. \emph{Inset}--Image of atoms in the science cell confirming transport. In-vacuum electrodes are highlighted by scattered light.}
\end{figure*}

We now verify that our new BDM cooling process will not limit realistic timescales for optical transport in upcoming continuous computing architectures (Fig.~3b). To study this we again image atoms in the lattice after a 50 ms dark time. We vary the BDM duration and fit an exponential saturation curve to find the characteristic time $\tau_{BDM}$ required to optimize optical trapping in the lattice. We measure  $\tau_{BDM}$=6.9(7) ms, comparable to optical transport times, the fastest natural timescale in this architecture.

\textit{Optical Transport}--Our motivation for demonstrating a Type-II BDM in Cs is to enable a static magnetic field architecture. Having shown sub-Doppler cooling and optical trapping with static magnetic fields, we now move to optical transport. After extinguishing all resonant light after the BDM, we apply a frequency ramp to one of the AOMs in the retro path (Fig. 4a), applying a linear acceleration of $\approx30$g before reversing the acceleration to bring atoms back to rest. With this hardware-limited acceleration profile, the 17 cm transport distance between MOT and the center of the science chamber takes 50 ms, a time we will reduce with upgraded control hardware.  To confirm atom transport, we set up a temporary vertical imaging path and image transported atoms within the science cell (Fig. 4b), verifying delivery of our Cs ensemble to the science cell, with the applied magnetic fields remaining constant during transport. 

A key figure of merit for continuous operation is the atomic flux delivered to the science cell. To evaluate our flux, we apply a second transport stage, returning our transported sample to the source cell where we image it with the Type-I MOT beams. We calibrate our atom number via camera and photodiode signal, with both agreeing to better than 20\%. For cycle rate we exclude the time for the return trip of the atoms (50 ms). As shown in Fig. 4b, we then vary the Type-I MOT loading time and thus cycle time of the experiment, measuring the round trip atom number. We indeed find that our transported atom number shows an exponential saturation time constant of $\tau =301(110)$  ms, verifying that our dynamics are dictated by the source cell vacuum lifetime. For our exponential saturation curve the optimal flux is obtained at $\tau$, and from our fit we find a flux of $3.0(4)\times10^6$ atoms/s. This demonstrates sufficient atomic flux for continuous operation with large scale qubit arrays. Further, this understanding of transport flux in relation to source cell background pressure supports tuning of the Cs background pressure to match desired operational cycle times. 

 \textit{Summary}--We have demonstrated magneto-optical trapping, sub-Doppler laser cooling, optical trapping, and optical transport of Cs with a single, static magnetic field configuration, establishing a new architecture for continuously operating alkali platforms. By demonstrating the first Type-II BDM in Cs, we eliminated the obstacle of sub-Doppler cooling in a static magnetic field. Finally, we showed that our transport provides millions of atoms per second to a science cell 17 cm away, providing a viable pathway toward qubit arrays containing $10^4-10^5$ atoms in next-generation quantum computers. This technique may enable realization of a core-shell MOT as has been realized in Yb~\cite{lee2015core}, further enhancing flux while suppressing scatter. Beyond this, we anticipate this architecture will work in Rb, where the first BDM was demonstrated. With recent demonstrations of interspecies Cs-Rb gates~\cite{miles2026qubit} and formation of RbCs molecules via tweezer merging~\cite{ruttley2023formation}, this new architecture offers exciting prospects for emerging frontiers in alkali-based quantum platforms.

%~\cite{SMcite}. 

\textit{Acknowledgments.}--We thank members of the broader Infleqtion team, notably A. Woolverton, A. Scott, M. Bedalov, M. Saffman, M. Gillette, and E. Salim.

\textit{Note added}--Recently we became aware of related BDM work in Cs~\cite{Hussein2026TypeII}.

\bibliography{references.bib}

@article{chen2022continuous,
  title={Continuous {B}ose--{E}instein condensation},
  author={Chen, Chun-Chia and Gonz{\'a}lez Escudero, Rodrigo and Min{\'a}{\v{r}}, Ji{\v{r}}{\'\i} and Pasquiou, Benjamin and Bennetts, Shayne and Schreck, Florian},
  journal={Nature},
  volume={606},
  number={7915},
  pages={683--687},
  year={2022},
  publisher={Nature Publishing Group UK London}
}

@article{chiu2025continuous,
  title={Continuous operation of a coherent 3,000-qubit system},
  author={Chiu, Neng-Chun and Trapp, Elias C and Guo, Jinen and Abobeih, Mohamed H and Stewart, Luke M and Hollerith, Simon and Stroganov, Pavel L and Kalinowski, Marcin and Geim, Alexandra A and Evered, Simon J and others},
  journal={Nature},
  volume={646},
  number={8087},
  pages={1075--1080},
  year={2025},
  publisher={Nature Publishing Group UK London}
}

@article{norcia2024iterative,
  title={Iterative assembly of 171 {Yb} atom arrays with cavity-enhanced optical lattices},
  author={Norcia, Matthew A and Kim, Hyemin and Cairncross, William B and Stone, M and Ryou, A and Jaffe, M and Brown, MO and Barnes, K and Battaglino, P and Bohdanowicz, TC and others},
  journal={PRX Quantum},
  volume={5},
  number={3},
  pages={030316},
  year={2024},
  publisher={APS}
}

@article{dalibard1989laser,
  title={Laser cooling below the Doppler limit by polarization gradients: simple theoretical models},
  author={Dalibard, Jean and Cohen-Tannoudji, Claude},
  journal={Journal of the Optical Society of America B},
  volume={6},
  number={11},
  pages={2023--2045},
  year={1989},
  publisher={Optical Society of America}
}

@article{tarbutt2015magneto,
  title={Magneto-optical trapping forces for atoms and molecules with complex level structures},
  author={Tarbutt, MR},
  journal={New Journal of Physics},
  volume={17},
  number={1},
  pages={015007},
  year={2015},
  publisher={IOP Publishing}
}

@article{devlin2016three,
  title={Three-dimensional Doppler, polarization-gradient, and magneto-optical forces for atoms and molecules with dark states},
  author={Devlin, JA and Tarbutt, MR},
  journal={New Journal of Physics},
  volume={18},
  number={12},
  pages={123017},
  year={2016},
  publisher={IOP Publishing}
}

@article{jarvis2018blue,
  title={Blue-detuned magneto-optical trap},
  author={Jarvis, Kyle N and Devlin, JA and Wall, TE and Sauer, BE and Tarbutt, MR},
  journal={Physical review letters},
  volume={120},
  number={8},
  pages={083201},
  year={2018},
  publisher={APS}
}

@article{burau2023blue,
  title={Blue-detuned magneto-optical trap of molecules},
  author={Burau, Justin J and Aggarwal, Parul and Mehling, Kameron and Ye, Jun},
  journal={Physical Review Letters},
  volume={130},
  number={19},
  pages={193401},
  year={2023},
  publisher={APS}
}

@article{li2024blue,
  title={Blue-detuned magneto-optical trap of {CaF} molecules},
  author={Li, Samuel J and Holland, Connor M and Lu, Yukai and Cheuk, Lawrence W},
  journal={Physical review letters},
  volume={132},
  number={23},
  pages={233402},
  year={2024},
  publisher={APS}
}

@article{petrich1994behavior,
  title={Behavior of atoms in a compressed magneto-optical trap},
  author={Petrich, Wolfgang and Anderson, Michael H and Ensher, Jason R and Cornell, Eric A},
  journal={Journal of the optical society of America B},
  volume={11},
  number={8},
  pages={1332--1335},
  year={1994},
  publisher={Optical Society of America}
}

@article{ketterle1999making,
  title={Making, probing and understanding {Bose-Einstein} condensates},
  author={Ketterle, Wolfgang and Durfee, Dallin S and Stamper-Kurn, DM},
  journal={arXiv preprint cond-mat/9904034},
  year={1999}
}

@book{metcalf1999laser,
  title={Laser cooling and trapping},
  author={Metcalf, Harold J and Van der Straten, Peter},
  year={1999},
  publisher={Springer Science \& Business Media}
}

@article{monroe1990very,
  title={Very cold trapped atoms in a vapor cell},
  author={Monroe, C and Swann, W and Robinson, H and Wieman, C},
  journal={Physical Review Letters},
  volume={65},
  number={13},
  pages={1571},
  year={1990},
  publisher={APS}
}

@incollection{grimm2000optical,
  title={Optical dipole traps for neutral atoms},
  author={Grimm, Rudolf and Weidem{\"u}ller, Matthias and Ovchinnikov, Yurii B},
  booktitle={Advances in atomic, molecular, and optical physics},
  volume={42},
  pages={95--170},
  year={2000},
  publisher={Elsevier}
}

@article{klostermann2022fast,
  title={Fast long-distance transport of cold cesium atoms},
  author={Klostermann, Till and Cabrera, Cesar R and von Raven, Hendrik and Wienand, Julian F and Schweizer, Christian and Bloch, Immanuel and Aidelsburger, Monika},
  journal={Physical Review A},
  volume={105},
  number={4},
  pages={043319},
  year={2022},
  publisher={APS}
}

@article{arpornthip2012vacuum,
  title={Vacuum-pressure measurement using a magneto-optical trap},
  author={Arpornthip, T and Sackett, CA and Hughes, KJ},
  journal={Physical Review A—Atomic, Molecular, and Optical Physics},
  volume={85},
  number={3},
  pages={033420},
  year={2012},
  publisher={APS}
}

@article{lee2015core,
  title={Core-shell magneto-optical trap for alkaline-earth-metal-like atoms},
  author={Lee, Jeongwon and Lee, Jae Hoon and Noh, Jiho and Mun, Jongchul},
  journal={Physical Review A},
  volume={91},
  number={5},
  pages={053405},
  year={2015},
  publisher={APS}
}

@article{meiser2009prospects,
  title={Prospects for a millihertz-linewidth laser},
  author={Meiser, Dominic and Ye, Jun and Carlson, DR and Holland, MJ},
  journal={Physical review letters},
  volume={102},
  number={16},
  pages={163601},
  year={2009},
  publisher={APS}
}

@article{holland1996theory,
  title={Theory of an atom laser},
  author={Holland, M and Burnett, K and Gardiner, C and Cirac, JI and Zoller, P},
  journal={Physical Review A},
  volume={54},
  number={3},
  pages={R1757},
  year={1996},
  publisher={APS}
}

@inproceedings{dick1989local,
  title={Local oscillator induced instabilities in trapped ion frequency standards},
  author={Dick, G John},
  booktitle={Proceedings of the 19th Annual Precise Time and Time Interval Systems and Applications Meeting},
  pages={133--147},
  year={1989}
}

@article{ruttley2023formation,
  title={Formation of ultracold molecules by merging optical tweezers},
  author={Ruttley, Daniel K and Guttridge, Alexander and Spence, Stefan and Bird, Robert C and Le Sueur, C Ruth and Hutson, Jeremy M and Cornish, Simon L},
  journal={Physical Review Letters},
  volume={130},
  number={22},
  pages={223401},
  year={2023},
  publisher={APS}
}

@article{miles2026qubit,
  title={Qubit syndrome measurements with a high fidelity {Rb-Cs} Rydberg gate},
  author={Miles, J and Lichtman, MT and Scott, AM and Scott, J and Norrell, SA and Bedalov, MJ and Belknap, DA and Cole, DC and Eubanks, SY and Gillette, M and others},
  journal={arXiv preprint arXiv:2603.13492},
  year={2026}
}

@article{li2024high,
  title={High-rate and high-fidelity modular interconnects between neutral atom quantum processors},
  author={Li, Yiyi and Thompson, Jeff D},
  journal={PRX Quantum},
  volume={5},
  number={2},
  pages={020363},
  year={2024},
  publisher={APS}
}

@inproceedings{aharonov1997fault,
  title={Fault-tolerant quantum computation with constant error},
  author={Aharonov, Dorit and Ben-Or, Michael},
  booktitle={Proceedings of the twenty-ninth annual ACM symposium on Theory of computing},
  pages={176--188},
  year={1997}
}

@article{kitaev1997quantum,
  title={Quantum computations: algorithms and error correction},
  author={Kitaev, A Yu},
  journal={Russian Mathematical Surveys},
  volume={52},
  number={6},
  pages={1191--1249},
  year={1997}
}

@article{jauregui2001anharmonic,
  title={Anharmonic parametric excitation in optical lattices},
  author={J{\'a}uregui, R and Poli, Nicola and Roati, G and Modugno, Giovanni},
  journal={Physical Review A},
  volume={64},
  number={3},
  pages={033403},
  year={2001},
  publisher={APS}
}

@article{kwolek2022continuous,
  title={Continuous sub-Doppler-cooled atomic beam interferometer for inertial sensing},
  author={Kwolek, JM and Black, AT},
  journal={Physical Review Applied},
  volume={17},
  number={2},
  pages={024061},
  year={2022},
  publisher={APS}
}

@misc{Hussein2026TypeII,
  author       = {Omar Hussein and Forouzan Forouhar-Manesh and Paul Del Franco and Megan Byres and Max Jones and Andrew Lagno and Addison Okell and Alan Jamison},
  title        = {Type-{II} Blue-detuned MOT for Cesium and Lithium},
  year         = {2026},
  note         = {Conference abstract, DAMOP 2026},
}

\section{Appendix A: Multiplexing Type-I MOT and Type-II BDM Paths}

The handedness of the Type-II BDM light is opposite to the Type-I light due to a combination of blue-detuning, excited state total angular momentum ($F'$), and excited state magnetic sensitivity~\cite{devlin2016three}. To deliver the BDM light, we phase-lock a second 852 nm laser to the Type-I laser with a variable frequency offset. We then overlap the Type-II BDM light onto the Type-I light in free space using a combination of polarizing beam-splitter (PBS) and 50:50 non-polarizing beam-splitter (NPBs), efficiently utilizing all light (Fig.~5). Each MOT arm thus has Type-I and Type-II light with opposite linear polarization. This avoids changing polarizations between the MOT stages as previously done in Rb~\cite{jarvis2018blue}, requiring only basic intensity control via AOMs.

 \begin{figure}[h]
	\includegraphics[trim={0 0.0cm 0 0.0cm},clip,width=.49\textwidth]{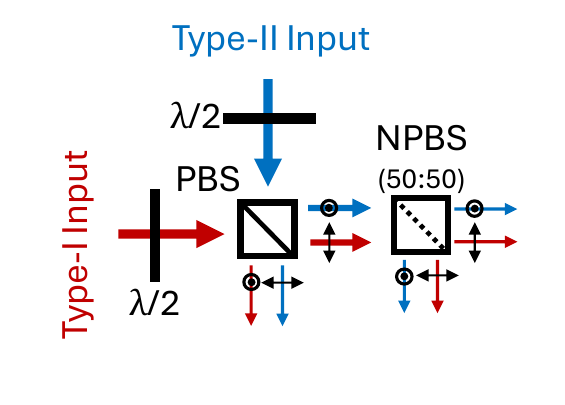}
	\caption{\label{5} Diagram showing free space combination and splitting of Type-I and Type-II lasers. Half-waveplates ($\lambda/2$) allow for setting the fraction of light that is transmitted (reflected) from the initial PBS, allowing 33\% of each incoming beam to be sent to the first MOT arm. A subsequent 50:50 NPBS then splits the remaining 66\% of the light in half, sending 33\% of each laser's total power into the other two MOT arms.}
	\label{figure_5}
\end{figure}

\end{document}